\documentclass[12pt]{article}
\pdfoutput=1
\usepackage{scicite}
\usepackage{times}
\usepackage{graphicx}
\usepackage{hyperref}

\usepackage{amssymb,amsmath}
\usepackage{deluxetable}

\newenvironment{sciabstract}{%
\begin{quote} \bf}
{\end{quote}}

\newcounter{lastnote}

\title{A Radio Pulsar/X-ray Binary Link}
\author{Anne M. Archibald$^{1\ast}$,  
Ingrid H. Stairs$^{2,3,4}$,
Scott M. Ransom$^5$,
Victoria M. Kaspi$^1$,\\
Vladislav I. Kondratiev$^{6,5,7}$,
Duncan R. Lorimer$^{6,8}$,
Maura A. McLaughlin$^{6,8,9}$,\\
Jason Boyles$^{6,8}$,
Jason W. T. Hessels$^{10,11}$,
Ryan Lynch$^{12}$,
Joeri van Leeuwen$^{10,11}$,\\
Mallory S. E. Roberts$^{13}$,
Frederick Jenet$^{14}$,
David J. Champion$^{3}$,\\
Rachel Rosen$^8$, 
Brad N. Barlow$^{15}$,
Bart H. Dunlap$^{15}$ \&
Ronald A. Remillard$^{16}$\\
\\
\footnotesize{$^{1}$Department of Physics, McGill University,}\\[-7pt]
\footnotesize{3600 rue University, Montr\'eal, QC, Canada H3A 2T8}\\[-7pt]
\footnotesize{$^{2}$Department of Physics and Astronomy, University of British Columbia,}\\[-7pt]
\footnotesize{6224 Agricultural Road, Vancouver, BC V6T 1Z1}\\[-7pt]
\footnotesize{$^{3}$Australia Telescope National Facility, Commonwealth Scientific and Industrial Research Organisation}\\[-7pt]
\footnotesize{PO Box 76, Epping NSW 1710, Australia}\\[-7pt]
\footnotesize{$^{4}$Centre for Astrophysics \& Supercomputing, Swinburne University of Technology}\\[-7pt]
\footnotesize{Mail 39 PO Box 218 Hawthorn Vic 3122 Australia}\\[-7pt]
\footnotesize{$^{5}$National Radio Astronomy Observatory}\\[-7pt]
\footnotesize{520 Edgemont Rd., Charlottesville, VA  USA  22901}\\[-7pt]
\footnotesize{$^{6}$Department of Physics, West Virginia University}\\[-7pt]
\footnotesize{210 Hodges Hall, Morgantown, WV 26506}\\[-7pt]
\footnotesize{$^{7}$Astro Space Center of the Lebedev Physical Institute}\\[-7pt]
\footnotesize{Profsoyuznaya str. 84/32, Moscow 117997, Russia}\\[-7pt]
\footnotesize{$^{8}$National Radio Astronomy Observatory}\\[-7pt]
\footnotesize{Green Bank, WV 24944}\\[-7pt]
\footnotesize{$^{9}$Alfred P. Sloan Fellow}\\[-7pt]
\footnotesize{$^{10}$Netherlands Institute for Radio Astronomy (ASTRON)}\\[-7pt]
\footnotesize{Postbus 2, 7990 AA Dwingeloo, The Netherlands}\\[-7pt]
\footnotesize{$^{11}$Astronomical Institute ``Anton Pannekoek'', University of Amsterdam}\\[-7pt]
\footnotesize{Kruislaan 403, 1098 SJ Amsterdam, The Netherlands}\\[-7pt]
\footnotesize{$^{12}$Department of Astronomy, University of Virginia}\\[-7pt]
\footnotesize{Charlottesville, VA 22904-4325}\\[-7pt]
\footnotesize{$^{13}$Eureka Scientific, Inc.}\\[-7pt]
\footnotesize{2452 Delmer St. Suite 100, Oakland, CA 94602-3017}\\[-7pt]
\footnotesize{$^{14}$Department of Physics and Astronomy, University of Texas at Brownsville}\\[-7pt]
\footnotesize{80 Fort Brown, Brownsville, TX 78520}\\[-7pt]
\footnotesize{$^{15}$Dept. of Physics \& Astronomy, University of North Carolina at Chapel Hill}\\[-7pt]
\footnotesize{CB 3255, Phillips Hall, Chapel Hill, NC 27599-3255}\\[-7pt]
\footnotesize{$^{16}$MIT Kavli Institute for Astrophysics and Space Research}\\[-7pt]
\footnotesize{77 Massachusetts Avenue, 37-287, Cambridge, MA 02139}\\[-7pt]
\\
\footnotesize{$^\ast$To whom correspondence should be addressed; E-mail: \texttt{aarchiba@physics.mcgill.ca}}
}

\date{}

\newcommand{\kpc}{\text{kpc}}

\newcommand{\degrees}{^\circ}
\newcommand{\arcmin}{\prime}
\newcommand{\arcsec}{{\prime\prime}}
\newcommand{\sun}{\odot}
\begin{document}

\maketitle

\clearpage
\begin{sciabstract}
Radio pulsars with millisecond spin periods are thought to have been spun up by transfer of matter and angular momentum from a low-mass companion star during an X-ray-emitting phase.  The spin periods of the neutron stars in several such low-mass X-ray binary (LMXB) systems have been shown to be in the millisecond regime, but no radio pulsations have been detected.  Here we report on detection and follow-up observations of a nearby radio millisecond pulsar (MSP) in a circular binary orbit with an optically identified companion star.  Optical observations indicate that an accretion disk was present in this system within the last decade. Our optical data show no evidence that one exists today, suggesting that the radio MSP has turned on after a recent LMXB phase.
\end{sciabstract}

% LMXBs and recycling. 
The fastest-spinning radio MSPs are thought to be formed in systems containing a neutron star (NS) and a low-mass ($\lesssim 1 M_\sun$) companion star \cite{bv91}.  Mass transfer occurs when matter overflows the companion's Roche lobe \cite{rochelobe}, forms an accretion disk around the NS, and eventually falls onto its surface, producing bright X-ray emission \cite{bv91,tv06}. 

% Evolves down to "period gap", companion becomes degenerate, period begins increasing again; why is there accretion when the companion is degenerate?
Radio emission that would otherwise be produced by a rapidly rotating magnetic NS is thought to be quenched during active accretion by the presence of ionized material within the pulsar's light cylinder.  However, for a sufficiently low accretion rate, infalling material may be halted outside the light cylinder by magnetic pressure, presumably allowing the MSP's radio emission to turn on.  At this point, the MSP's electromagnetic emission and particle wind should irradiate the disk and companion, driving mass out of the system. This activation as a radio pulsar, a possible end to the accretion phase, is poorly understood, in part because no radio MSP has yet been observed to turn on in such a system, although there is indirect evidence that this may occur \cite{bdd+03,hpc+08,dhtj08}.

% The source and its history
The radio source FIRST J102347.67+003841.2 (hereafter J1023) appears optically as a $V\sim 17.5$ magnitude star with a mid-G (solar-type) spectrum and mild 0.198-day orbital variability \cite{ta05}. However, in observations from May 2000 to Dec. 2001, J1023 had a blue spectrum with prominent emission lines \cite{bwbo02,sfs+03} (absent in quiescence) and exhibited rapid flickering by $\sim 1$ magnitude \cite{bwbo02}. This optical behavior is typical of an accretion flow, and the double-peaked nature of the lines \cite{sfs+03} suggests an accretion disk specifically. This history led to the system's classification as an accreting white dwarf \cite{bwbo02} or NS \cite{ta05} binary. For a more detailed history see the Supporting Online Material (SOM). Since 2002 the system has shown no further sign of accretion, and our optical spectroscopic observations of Dec. 23 and 25, 2008 (see SOM) confirm the continued absence of emission lines. Thus J1023 is currently in a quiescent state. 

% Discovery of radio pulsations.
We found PSR~J1023+0038, a bright MSP with a spin period of $1.69\;\text{ms}$ at a dispersion measure (DM) of $14.33\;\text{pc}\;\text{cm}^{-3}$, as part of a 350-MHz pulsar survey carried out with the Robert C. Byrd Green Bank Telescope (GBT) in 2007 (see SOM for details).  

% Inferred binary data.
Phase-coherent timing of the radio pulsations at the GBT, the Arecibo Observatory, and Parkes Observatory has allowed high-precision measurements of the binary parameters (see Table~\ref{tab:params} and SOM for details). Together with the much smaller position uncertainties implied by the telescope beam sizes of the timing observations, these binary parameters unambiguously place this pulsar in the J1023 system, confirming the prediction of \cite{ta05} that J1023 has a NS primary.

Combining these parameters with optical radial velocity measurements \cite{ta05}, we find that the pulsar is a factor of $7.1\pm 0.1$ more massive than its companion.  The determination of the individual masses depends only on the inclination angle $i$ of the orbital plane from the plane of the sky.  The pulsar's Keplerian mass function $(m_c \sin i)^3/(m_p+m_c)^2$ is $1.1\times 10^{-3} \; M_\sun$, where $m_c$ and $m_p$ are the masses of the companion and the pulsar, respectively.  For a NS mass in the range $\sim 1.0$ to $\sim 3.0~M_\sun$ \cite{lp04}, $i$ is restricted to the range $\sim 53\degrees$ to $\sim 34\degrees$, and the companion mass is restricted to $\sim 0.14$ to $\sim 0.42 M_\sun$.

Stellar models, based on the assumption (reasonable in light of the evidence for a recent accretion flow) that the companion is filling its Roche lobe, show that it is possible for the companion star to exhibit the observed orbital modulation of color, magnitude, and radial velocity, if the primary has an isotropic luminosity of $\sim 2~L_\sun$ \cite{ta05}. These models also imply $i<55\degrees$, which is compatible with the range obtained from theoretical NS mass limits.
Similar models account for the light curve modulation during quiescence of the low-mass companion to SAX J1808.4$-$3658 \cite{dhtj08}, an accreting X-ray pulsar in a $2\;\text{hr}$ orbit (e.g. \citen{wk98,cm98}), and which has been suspected of harboring an MSP when in quiescence \cite{bdd+03,hjw+09,hpc+08}.  

The brightness, radial velocities, and presumed Roche lobe filling companion of J1023 imply that the distance to the system is $0.9~\kpc/\sin i$; for the allowed range of $i$, the distance is between $\sim 1.1$ and $\sim 1.6~\kpc$.  A pulsar mass of $1.4 M_\sun$ implies a distance of $1.3~\kpc$. For comparison, given the observed DM of $14.325\;\text{pc}\;\text{cm}^{-3}$ the standard Galactic free-electron density model \cite{cl02} predicts a distance of $0.6~\text{kpc}$, consistent with our range given its known uncertainties.

Ordinarily one can estimate a pulsar's spin-down luminosity from its observed spin-down $\dot P$, but for MSPs this is subject to contamination by several factors, including accelerations of J1023 and the solar system barycenter in the Galactic gravitational potential.  From the measured optical proper motion $\mu$ of J1023 (Table~\ref{tab:params})
we expect a significant positive contribution to $\dot{P} = Pd\mu^2/c \simeq 1.8 \times 10^{-21} (d/1.3 \; \kpc)$ due to the Shklovskii effect \cite{shkl70}. While in some binary systems the accelerations can be constrained independently using orbital period variations,
in J1023 we observe very large orbital period variability (binary period derivative $\sim 3\times 10^{-10}$), almost certainly due to classical tidal torquing from gas motion in the extended envelope of the companion star.
The near-zero eccentricity of the pulsar's orbit (Table~\ref{tab:params}) also suggests that the system has undergone tidal circularization.

\begin{figure}
\begin{center}
\includegraphics[width=0.7\textwidth]{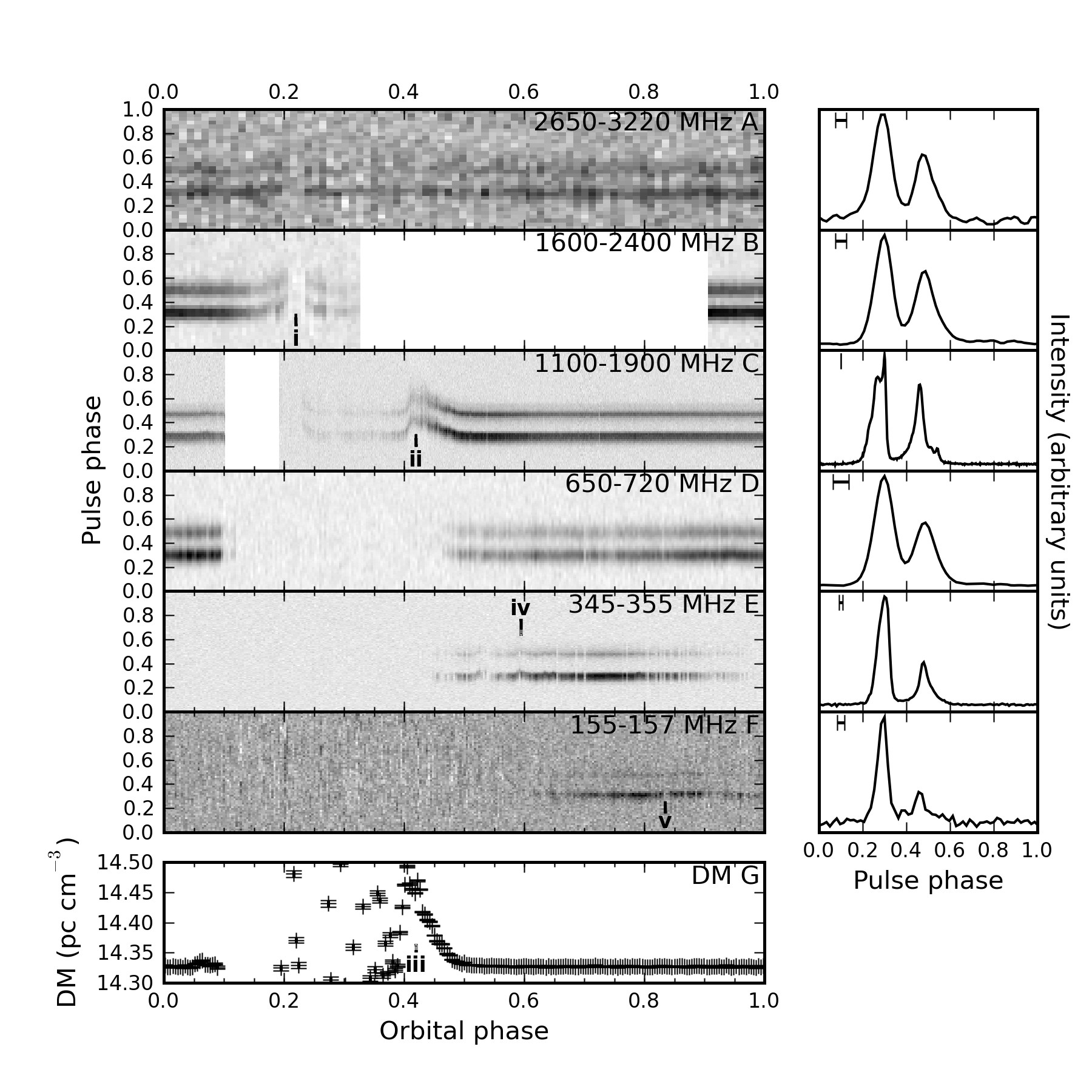}
\end{center}
\caption{\label{fig:orbplot}
Frequency dependence of eclipses, DM variations, and pulse profiles. \textbf{(A--F)} show flux density as a function of pulse phase and orbital phase at $3000$, $2000$, $1600$, $700$, $350$, and $156\;\text{MHz}$, respectively. \textbf{(G)} shows DM, as estimated from the $1600\;\text{MHz}$ observations (see SOM for details). 
Orbital phase is defined to be zero at the pulsar's ascending node, so that the companion passes closest to our line of sight at orbital phase $0.25$. 
To the right of \textbf{(A)} through \textbf{(F)} are pulse profiles at the respective bands; the instrumental smearing time is indicated with a horizontal bar in each panel. The exception is that the profile \textbf{(C)} is instead based on a $1410~\text{MHz}$ observation with higher time resolution.
Note that the eclipse is nearly absent at $3000\;\text{MHz}$, but is longer at $700\;\text{MHz}$ than at $1600\;\text{MHz}$. At low frequencies, random short eclipses, indicated by \textbf{(iv)} and \textbf{(v)}, are also visible.
Pulse phase is as predicted by a phase-coherent timing solution, so that large short-term timing variations are visible as vertical motions of the pulse peak. Note in particular the pulse arrival time variations at eclipse ingress \textbf{(i)} and egress \textbf{(ii)}; the variation at egress appears to be due to the substantial DM variation \textbf{(iii)} visible in \textbf{(G)}.
}
\end{figure}

% Timing issues. Eclipses/gas in system.
This picture of a Roche-lobe-filling companion is supported by evidence for the presence of ionized material in the system.
Our broad-band radio observations reveal substantial DM variations at certain orbital phases (Fig.~\ref{fig:orbplot}~G) as well as smaller DM variations on time scales of minutes throughout the orbit.  Moreover, we observe regular eclipses ranging from very brief at $3000~\text{MHz}$ to most of the orbit in our $150~\text{MHz}$ observations with the Westerbork Synthesis Radio Telescope (Fig.~\ref{fig:orbplot}~A--F). We also see very brief eclipses at all orbital phases (Fig.~\ref{fig:orbplot}~E in particular).  Given the established range of $i$, the line-of-sight between the pulsar and the Earth will not intersect the Roche lobe of the companion at any point in the orbit.  If we assume that during the DM variations near orbital phase 0.4 (Fig.~\ref{fig:orbplot} ii and iii) the obstructing plasma is uniform, has a plasma frequency of $700\;\text{MHz}$ (based on the loss of the $700\;\text{MHz}$ signal at this phase), and contributes the $0.15\;\text{pc}\;\text{cm}^{-3}$ excess DM, we predict a thickness of $3\times 10^4\;\text{km}$ (thin compared to the companion size of $\sim 3\times 10^5\;\text{km}$). This suggests that a thin but dense layer of material, perhaps a shock due to the pulsar wind meeting either winds from the companion or material overflowing the companion's Roche lobe, crosses the line of sight at this orbital phase.

Similar eclipses and DM variations have been seen in the Galactic ``black widow'' systems, tight NS binary systems in the Galactic field with companions an order of magnitude less massive than that of J1023 (e.g. \citen{fgld88}). Such eclipses have also been seen in several radio MSPs in globular clusters (e.g. \citen{dpm+01,sbl+96}). In both cases they have been attributed to the presence of gas flowing out from the irradiated companion, as is probably the case for J1023.
However J1023 differs from all other known radio MSPs in that there is evidence for a recent accretion disk.  

% Why we think it's accreting.
For an accreting NS the accretion luminosity is proportional to the accretion rate, which is unknown for J1023. But if the matter is to reach the NS surface, it must overcome magnetic pressure at the Keplerian corotation radius, so there is a minimum accretion rate and hence a minimum luminosity for a given NS magnetic field \cite{ccm+98}. For a magnetic field equal to our timing-derived upper limit (Table~\ref{tab:params}) this minimum luminosity for J1023 is $1.1\times 10^{37}~\text{erg}~\text{s}^{-1}$ (using formulae from \citen{ccm+98}). Such a high luminosity can be clearly ruled out by our analysis of archival \emph{Rossi X-ray Timing Explorer} all-sky monitor data (see SOM for details): the average annual X-ray luminosity for each year 1996--2008 was less than $4.8\times 10^{33}\;\text{erg}\;\text{s}^{-1} (d/1.3\;\kpc)^2$. The variable nature of LMXB X-ray emission cannot explain this low luminosity: on Feb 1, 2001, when an optical emission line spectrum was observed \cite{sfs+03}, the average flux was less than $2.4\times 10^{34}\;\text{erg}\;\text{s}^{-1} (d/1.3\;\kpc)^2$. If accretion occurred during the active phase, the annual upper limit would imply that the magnetic field of the NS in J1023 was less than $6\times 10^6~\text{G}$, smaller than that of any known MSP. 

It therefore seems more likely that infalling matter did not reach the NS surface, but instead underwent what is known as ``propeller-mode accretion'' \cite{is75}: infalling matter entered the light cylinder but was stopped by magnetic pressure outside the corotation radius, which prevented it from falling further inward.
This process also has a minimum accretion rate and minimum luminosity: if the infalling matter does not reach the light cylinder, the radio pulsar mechanism will presumably become active. If this occurs no stable balance exists between outward radiation and wind pressure and ram pressure of infalling material, and the disk will be cleared from the system \cite{ccm+98}. The minimum luminosity for propeller-mode accretion is substantially lower than that for standard accretion: our timing-derived upper limit on the magnetic field (Table~\ref{tab:params}) implies a minimum luminosity of just $2.7\times 10^{32}~\text{erg}~\text{s}^{-1}$ (using formulae from \citen{ccm+98}). Thus it seems likely that during J1023's active phase, the mass transfer rate was high enough for propeller-mode accretion, but not high enough for material to reach the NS surface. 

In this scenario, a small drop in the mass transfer rate would clear the disk from the system and return J1023 to its current quiescent state. Overflowing matter from the companion would then encounter a shock near the inner Lagrange point and be carried out of the system \cite{ccm+98}, possibly explaining the observed variable hard power-law X-ray spectrum of J1023 \cite{hsc+06} and the presence of gas in the system.
Such shocks have been suggested to explain the variable power-law X-ray emission in 47~Tuc~W \cite{bgv05} and PSR~J1740$-$5340 in NGC 6397 \cite{gch+02}, which also have low-mass companions, and have previously been compared with SAX J1808.4$-$3658, though with no evidence for the recent presence of an accretion disk.  

Despite the apparent resemblance of J1023 to 47~Tuc~W and PSR~J1740$-$5340, the latter two systems (and indeed all other currently known eclipsing pulsars with several-tenths $M_\sun$ companions) reside in globular clusters, and are likely to have acquired their current companions in exchange interactions after the pulsars were ``recycled'' \cite{cr05}.  J1023 is the only known highly recycled (the fifth fastest known) MSP in the field of the Galaxy with both a non-degenerate companion and an orbit that has been circularized through tidal interactions.  
A globular cluster origin for J1023 is extremely unlikely due to its large distance from the nearest globular cluster as well as from the Galactic bulge.
The evidence points to J1023's having been recycled by its current companion, which has not yet completed the transformation to a white dwarf. 

The observed transition of J1023 suggests that it is in a bistable state: for certain rates of Roche lobe overflow, if the radio pulsar mechanism is quenched, propeller-mode accretion can occur, but if the radio pulsar mechanism is active, that same mass accretion rate cannot overcome the radiation pressure and no accretion occurs. 
Should the mass transfer rate of J1023 rise sufficiently, then, it may enter another LMXB phase: a disk will form, the radio emission may be quenched, and the X-ray luminosity may increase dramatically, due to either propeller-mode accretion, with a net spin-down of the pulsar, or even brighter accretion onto the surface, with a net spin-up.

% table is here so that references get numbered appropriately
\begin{table}
\begin{center}
\begin{tabular}{ll}
\hline
Parameter & Value \\
\hline
Right ascension ($\alpha$; J2000) & $10^{\textrm{h}} 23^{\textrm{m}} 47.687(3)^{\textrm{s}}$ \\
Declination ($\delta$; J2000) & $00\degrees 38^\arcmin 41.15(7)^\arcsec $ \\
Proper motion in $\alpha$ ($\mu_\alpha$) & $10(1)\;\text{mas}\;\text{yr}^{-1}$ \\
Proper motion in $\delta$ ($\mu_\delta$) & $-16(2)\;\text{mas}\;\text{yr}^{-1}$ \\
Epoch (MJD) & $54802$ \\
\hline
Dispersion Measure & $14.325(10)\;\text{pc}\;\text{cm}^{-3}$ \\
Pulsar period ($P$) & $1.6879874440059(4)\;\text{ms}$ \\
Pulsar period derivative  ($\dot P$) & $1.2(8) \times 10^{-20}$ \\
Orbital period ($P_b$) & $0.1980962019(6)\;\text{d}$ \\
Orbital period derivative ($\dot P_b$) & $2.5(4)\times 10^{-10}$ \\
Orbital period second derivative ($\ddot P_b$) & $-5.21(14)\times 10^{-11}\;\text{s}^{-1}$ \\
Time of ascending node (MJD) & $54801.97065348(9)$ \\
$a\sin i$ & $0.3433494(3)\;\text{lt-s}$ \\
Eccentricity & $\lesssim 2\times 10^{-5}$ \\
\hline
$1600\;\text{MHz}$ flux density & $\sim 14\;\text{mJy}$ \\
Spectral index ($\alpha$; $S\propto \nu^{-\alpha}$) & $\sim -2.8$ \\
Surface magnetic field & $< 3\times 10^8\;\text{G}$ \\
Spin-down luminosity & $< 3\times 10^{35}\;\text{erg}\;\text{s}^{-1}$ \\
\hline
\end{tabular}
\end{center}
\caption{\label{tab:params}
Pulsar parameters. Parameters listed in the top section are held fixed in the  timing fit that produces the parameters listed in the middle section (SOM). 
The position and proper motion values are from the USNO NOMAD optical catalog \cite{zml+04}.  
With the exceptions of DM, $P$, and eccentricity, uncertainties on timing quantities are twice the formal $1\sigma$ errors returned by the pulsar timing analysis program. The DM of the source varies substantially: around orbital phase 0.3 it occasionally increases by $\sim 0.15\;\text{pc}\;\text{cm}^{-3}$, and we observe apparent orbit-to-orbit variations as large as $\sim 0.01\;\text{pc}\;\text{cm}^{-3}$.  
Our estimate of the eccentricity is approximate because DM variations limit the orbital coverage of usable timing data. Our estimate for $\dot P$ is approximate because it is highly covariant with pulsar position; it was estimated by fitting for $\dot P$ while varying the position according to the optical uncertainties listed above. We have not subtracted Shklovskii or galactic accelerations.
The flux density and spectral index are estimated using standard system temperature and gain values for the GBT and the Parkes telescope. The flux density is subject to substantial variability intrinsic to the pulsar (we have observed brightening by a factor $\sim 4$ within minutes), as well as interstellar scintillation; for this reason we estimated the spectral index using the simultaneous $600~\text{MHz}$ and $3000~\text{MHz}$ Parkes observations. 
Spin-down luminosity and magnetic field upper limits are calculated from the $3\sigma$ upper limit on $\dot P$ using the standard formulae $\dot E = 3.95\times 10^{31}\;\text{erg}\;\text{s}^{-1} (\dot P/10^{-15}) (P/1\;\text{s})^{-3}$ and $B=3.2\times 10^{19}\;\text{G}\;\sqrt{P\dot P}$. 
}
\end{table}

\clearpage
\appendix
\section*{Supporting Online Material}

\paragraph{History of observations of J1023:} 

% The source and its history.
The source FIRST J102347.67+003841 was detected as  a point source in 1998 August in the FIRST VLA 1.4~GHz continuum radio sky survey \cite{bwh95}. Three observations over the course of  a week showed that the flux density of J1023 varied from less than a  few mJy to roughly 7~mJy \cite{bwbo02}.
The source was also identified with a $V\sim 17.5$ magnitude star having a history of observations back to a 1952 Palomar Observatory Sky Survey observation \cite{zml+04}. In 1999 March, a Sloan Digital Sky Survey filter-photometry observation showed that this star had colors consistent with those of a G star \cite{sfs+03}.

However, observations at optical wavelengths showed a dramatic change: spectra taken in May 2000 \cite{bwbo02} were substantially bluer, and had prominent emission lines, leading to classification of J1023 as a cataclysmic variable. Further spectroscopic observations \cite{sfs+03,bwbo02} confirmed this spectrum, and revealed emission lines with double peaks and orbital radial velocity modulation as late as December 2001. Optical photometry \cite{bwbo02} showed flickering, varying by $\sim 1$~mag on timescales from ten seconds to one day.

Optical observations after December 2001 showed a return to the previous state: a May 2002 observation \cite{hsc+06} showed a G-like spectrum  without emission lines, and photometry taken in late 2002 and early 2003 \cite{wwp04} shows a smooth light curve with a modulation period of $0.198\;\text{days}$, having amplitude $\sim$0.3~mag in V, though with some low-level flickering.  Observations in 2004 \cite{ta05} found a smooth light curve consistent with that of 2003, though with  no flickering, and a spectrum matching that of a mid-G star.  Radial  velocity measurements showed a regular Doppler shift with $0.198\;\text{day}$ period and peak velocity $268 \pm 4\;\text{km}\;\text{s}^{-1}$ \cite{ta05}.  Light curve and color-variation modelling indicated that the primary was too massive and too luminous to be a white dwarf (given the observed red optical spectrum), so it  was suggested that the source was probably an LMXB with a NS primary, now in quiescence \cite{ta05}. This idea was supported by \textit{XMM-Newton} observations in 2004, which found a hard power-law X-ray spectrum \cite{hsc+06}.  

Further details of all these observations are compiled in Table~\ref{tab:oldobs}.

\paragraph{Our observations of J1023:} 
In mid-2007, we carried out a 350-MHz pulsar survey with the
Robert C. Byrd Green Bank Telescope (GBT) in West Virginia.  This survey
covered $\sim 12,000$ square degrees of sky in the declination range
$-21\degrees$ to $+26\degrees$, and used the Spigot pulsar
autocorrelation spectrometer \cite{kel+05} to synthesize a filterbank
sampling a $50\;\text{MHz}$ band with 2048 frequency channels every
81.92 $\mu$s.  Details of the survey and ongoing data analysis
procedures will be presented elsewhere.  In the data from June
28, 2007, we found PSR~J1023+0038, a bright MSP (mean flux density $\sim
75\;\text{mJy}$ in the $140\;\text{s}$ discovery observation) with a
spin period of $1.69\;\text{ms}$ at a dispersion measure (DM) of $14.
33\;\text{pc}\;\text{cm}^{-3}$. 

We obtained follow-up radio observations; all these observations are summarized in Table~\ref{tab:ourobs}. These observations used several different instruments and backends, in several different radio-frequency bands. Where possible, we observed simultaneously with both coherent dedispersion backends with online folding (for precise timing) and incoherent dedispersion backends with offline folding (for larger bandwidths and to retain single-pulse and transient information). For these observations we used the GBT, the Arecibo Observatory (AO) in Puerto Rico, the Westerbork Synthesis Radio Telescope (WSRT) in the Netherlands, and the Parkes telescope in Australia. 
The GBT Pulsar Spigot is an autocorrelation spectrometer \cite{kel+05}. The Green Bank Ultimate Pulsar Processing Instrument\footnote{\url{https://safe.nrao.edu/wiki/bin/view/CICADA/NGNPP}} (GUPPI) is a new digital filterbank for the GBT. The Wideband Arecibo Pulsar Processors (WAPPs) are a set of autocorrelation spectrometers usable on independent frequency bands \cite{dsk00}. The Parkes 10/50 cm dual-band receiver feeds a set of 1-bit analog filterbanks. 
The Arecibo Signal Processor (ASP) and the Green Bank Astronomical Signal Processor (GASP) are systems that use a computer cluster to coherently dedisperse \cite{hr75} and fold incoming data using the best available ephemeris. Further details of the operation of ASP and GASP and the reduction of their output data can be found in \cite{dem07} and \cite{fer08}. 
We used the Pulsar Machine II (PuMa-II) coherent dedispersion backend at the WSRT (an interferometer we operated in phased-array mode). The precise modes we used for each observation are described in the footnotes to Table~\ref{tab:ourobs}.

\begin{figure}
\begin{center}
\includegraphics[width=0.9\textwidth]{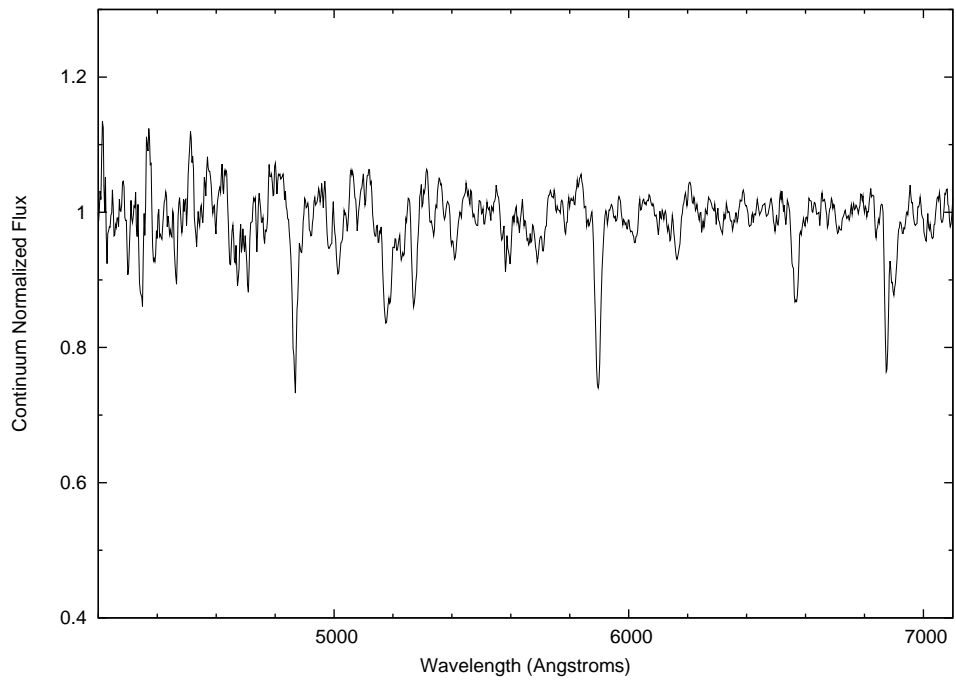}
\end{center}
\caption{\label{fig:spec}Optical spectrum for J1023, taken on Dec. 23 and 25, 2008. Flux was normalized by dividing by the best-fit continuum spectrum. Note the absence of emission lines; this spectrum is consistent with that observed in quiescence and not consistent with that observed during J1023's active phase.}
\end{figure}
In addition to the radio observations described above, we took an optical spectroscopic observation Dec. 23 and 25, 2008 with the Goodman high-throughput spectrograph on the Southern Astrophysical Research telescope. Since no flux calibration observation was taken, we produced a spectrum by averaging our seven 900-s exposures and dividing by the best-fit continuum. The resulting spectrum (Figure~\ref{fig:spec}) confirms the continued absence of emission lines, and hence quiescent state of the system.

\paragraph{X-ray upper limit:} In order to improve on the upper bound of \cite{hsc+06} on J1023's X-ray brightness during its active phase, we searched the archival \emph{Rossi X-ray Timing Explorer} All-Sky Monitor data for the known position of J1023.  This data covers the energy range 2--10 keV for all of the span 1996--2006 with the exception of from brief periods when the source passed close to the Sun. The average annual flux for each year 1996--2008 must be beneath the systematic upper limit of $2.4\times 10^{-11}\;\text{erg}\;\text{s}^{-1}\;\text{cm}^{-2}$ ($4.8\times 10^{33}\;\text{erg}\;\text{s}^{-1} (d/1.3\;\kpc)^2$). Moreover, since LMXBs are variable sources, we searched individual days as well, obtaining varying upper limits; in particular, the average X-ray flux during Feb 1, 2001, when an optical emission-line spectrum was observed \cite{sfs+03}, did not exceed $1.2\times 10^{-10}\;\text{erg}\;\text{s}^{-1}\;\text{cm}^{-2}$ (corresponding to a luminosity of $2.4\times 10^{34}\;\text{erg}\;\text{s}^{-1} (d/1.3\;\kpc)^2$).

\paragraph{Details of Main Text Figure~\ref{fig:orbplot}:}
The data shown in \textbf{(A)} and \textbf{(D)} were taken simultaneously with the Parkes $10$/$50\;\text{cm}$ dual-band receiver and analog filterbank on Nov. 24, 2008; the data shown in \textbf{(B)} were taken with the GBT Pulsar Spigot on Oct. 18, 2008; the data shown in \textbf{(C)} and \textbf{(G)} are the same $1600\;\text{MHz}$ observation taken with the GUPPI and the GBT Pulsar Spigot respectively on Nov. 1, 2008; and the data in \textbf{(E)} and \textbf{(F)} were taken with the Westerbork Synthesis Radio Telescope using the PuMa-II coherent dedispersion backend on Nov. 11, 2008 and Dec. 31, 2008, respectively. 

The pulse profile next to \textbf{(C)} is based on a $1400$-$\text{MHz}$ band observation taken at the Arecibo Observatory on Nov. 18, 2008 and coherently dedispersed with the ASP backend \cite{drb+04}, while the others are based on the same observations as the grayscale plots and generally suffer from poorer time resolution.  Smearing in the profiles obtained by coherent dedispersion is due to pulse drift and inaccurate DM values; the time and frequency subintegrations have been aligned with an improved ephemeris but smearing within each subintegration remains. 

\paragraph{Obtaining a timing solution:} From each observation of PSR~J1023+0038, we produced folded profiles for time and frequency subintegrations.  We then established a reference profile for each combination of band and backend, usually from the best observation available with that configuration, though we re-used the standard profile from the $1400\;\text{MHz}$ ASP observations for the $1388\;\text{MHz}$ and $1788\;\text{MHz}$ GASP observations. We produced a pulse time-of-arrival (TOA) for each subintegration (generally $30~\text{s}$ or $60~\text{s}$) by cross-correlating the reference profile with the folded profile for that subintegration \cite{tay92}. These arrival times are based on observatory time standards which were corrected to UTC using the GPS satellites. We fed these TOAs into the pulsar timing program \textsc{tempo}\footnote{\url{http://www.atnf.csiro.au/research/pulsar/tempo/}} to obtain a timing solution, allowing for an arbitrary time offset of each band/backend combination.  The solution we obtain is referenced to the DE405 solar system ephemeris.

The initial timing solution we obtained showed the timing effects of the DM variations we observed, as well as possibly some other systematics. We excluded any data that showed signs of short-term DM variations, including all data between orbital phases 0.15 and 0.55, as well as most low-frequency data. Specifically, the only data below $1~\text{GHz}$ we used was one pair of $800~\text{MHz}$ GBT observations separated by a day and one $350~\text{MHz}$ GBT observation. These data sets did not show signs of significant short-term DM variations, and the inter-instrument jump we assumed absorbed any effects of long-term DM variations.  To mitigate any remaining systematics, we artificially increased the uncertainties on each individual data set until it was fit with a reduced $\chi^2$ of approximately one. 

We have attached the resulting lists of pulse arrival times in TEMPO format to this supporting online material.

\paragraph{Estimating DM variations:} The global DM value we report is obtained from the Oct. 18, 2008 GUPPI observation. We produced TOAs for four frequency subbands for each time subintegration, aligning the folded data with a common template, and used the pulsar timing package \textsc{tempo} to fit for the DM that minimized the weighted residuals.  This process yields a very precise DM, but when we applied an analogous procedure to several other data sets, the results varied from $14.32$ to $14.33~\text{pc}~\text{cm}^{-3}$. We applied a similar process to the data taken at Arecibo with the WAPPs, obtaining a DM value for each time subintegration. These values showed the large DM excesses around orbital phase 0.3, but they also confirmed the variability of the average DM, giving values stable within an orbit but varying from orbit to orbit by as much as $0.01~\text{pc}~\text{cm}^{-3}$.

\begin{deluxetable}{lccl}
\tablecaption{\label{tab:oldobs} Published observations of J1023.}
\tabletypesize{\scriptsize}
\tablehead{\colhead{Date} &  \colhead{Type} & \colhead{Reference} & \colhead{Comment} }
\startdata
1992 Nov 7 & 436 MHz pulsar survey\tablenotemark{a} & \cite{lml+98} & No detection ($\lesssim 3~\text{mJy}$) \\
1998 Aug 3 & 1.4 GHz continuum\tablenotemark{b}  & \cite{bwbo02} & $<1.8~\text{mJy}$ \\
1998 Aug 8 & 1.4 GHz continuum\tablenotemark{b}  & \cite{bwbo02} & $<3.4~\text{mJy}$ \\
1998 Aug 10 & 1.4 GHz continuum\tablenotemark{b} & \cite{bwbo02} & $6.56~\text{mJy}$ \\
1999 Mar 22 & $u'$, $g'$, $r'$, $i'$, $z'$ photometry\tablenotemark{c} & \cite{aaa+08} & G star color \\
\hline
2000 May 6 & Spectroscopy\tablenotemark{d} & \cite{bwbo02} & Blue, emission lines \\
2000 May 9 & Spectroscopy\tablenotemark{e} & ESO archive & Blue, emission lines \\
2000 Nov 21 & Fast photometry\tablenotemark{f} & \cite{bwbo02} & Flickering by $\sim 1$ mag \\
2000 Nov 22 & Fast photometry\tablenotemark{f} & \cite{bwbo02} & Flickering by $\sim 1$ mag\\
2000 Dec 22 & Fast photometry\tablenotemark{f} & \cite{bwbo02} & Flickering by $\sim 1$ mag\\
2000 Dec 23 & Fast photometry\tablenotemark{f} & \cite{bwbo02} & Flickering by $\sim 1$ mag\\
2000 Dec 24 & Fast photometry\tablenotemark{f} & \cite{bwbo02} & Flickering by $\sim 1$ mag\\
2001 Feb 1 & Spectroscopy\tablenotemark{c} & \cite{aaa+08} & Blue, emission lines \\
2001 Dec 10 & Spectroscopy\tablenotemark{g} & \cite{sfs+03} & Blue, double-peaked emission lines \\
\hline
2002 May 11 & Spectropolarimetry\tablenotemark{h} & \cite{hsc+06} & G star spectrum, no significant polarization \\
2002 Dec 30 & Fast photometry\tablenotemark{i} & \cite{wwp04} & Smooth, low-amplitude flickering \\
2003 Oct 16 & 8.4 GHz continuum\tablenotemark{b}  & \cite{mg07} & No detection \\
2003 Jan 28 & Fast photometry\tablenotemark{j} & \cite{wwp04} & Smooth, low-amplitude flickering \\
2003 Jan 29 & Fast photometry\tablenotemark{j} & \cite{wwp04} & Smooth, low-amplitude flickering \\
2003 Jan 30 & Fast photometry\tablenotemark{j} & \cite{wwp04} & Smooth, low-amplitude flickering \\
2003 Jan 31 & Fast photometry\tablenotemark{j} & \cite{wwp04} & Smooth, low-amplitude flickering \\
2003 Feb 03 & Fast photometry\tablenotemark{j} & \cite{wwp04} & Smooth, low-amplitude flickering \\
2003 Feb 05 & Fast photometry\tablenotemark{j} & \cite{wwp04} & Smooth, low-amplitude flickering \\
2003 Feb 21 & Fast photometry\tablenotemark{j} & \cite{wwp04} & Smooth, low-amplitude flickering \\
2003 Feb 24 & Fast photometry\tablenotemark{j} & \cite{wwp04} & Smooth, low-amplitude flickering \\
2004 Feb 16 & Spectropolarimetry\tablenotemark{k} & \cite{hsc+06} & G star spectrum, no significant polarization \\
2004 May 12 & X-ray\tablenotemark{l} & \cite{hsc+06} & Hard power-law spectrum \\
2004 May 12 & X-ray\tablenotemark{m} & \cite{hsc+06} & Irregular, 60\% modulation \\
2004 May 12 & $B$ photometry\tablenotemark{n}  & \cite{hsc+06} & Smooth periodic 40\% modulation \\
2004 May 13 & Photometry\tablenotemark{o} & \cite{hsc+06} & Smooth periodic 30\% modulation \\
2004 May 23 & Spectroscopy\tablenotemark{p} & \cite{hsc+06} & \\
2003 Jan 31 & Spectroscopy\tablenotemark{p} & \cite{ta05} & \\
2004 Jan 18 & Spectroscopy\tablenotemark{p} & \cite{ta05} & \\
2004 Jan 19 & Spectroscopy\tablenotemark{p} & \cite{ta05} & \\
2004 Jan 20 & Spectroscopy\tablenotemark{p} & \cite{ta05} & \\
2004 Mar 8 &  Spectroscopy\tablenotemark{p} & \cite{ta05} & \\
2004 Mar 9 &  Spectroscopy\tablenotemark{p} & \cite{ta05} & \\
2004 Nov 18 & Spectroscopy\tablenotemark{p} & \cite{ta05} & \\
2004 Feb 29 & Filter photometry\tablenotemark{p} & \cite{ta05} & \\
2004 Feb 29 & Fast $I$ photometry\tablenotemark{p} & \cite{ta05} & \\
2004 Mar 1 &  Fast $I$ photometry\tablenotemark{p} & \cite{ta05} & \\
2004 May 12 & Fast $B$, $V$, $I$ photometry\tablenotemark{q}  & \cite{ta05} & \\
2004 May 16 & Fast $B$, $V$, $I$ photometry\tablenotemark{q} & \cite{ta05} & \\
2004 May 20 & Fast $B$, $V$, $I$ photometry\tablenotemark{q} & \cite{ta05} & \\
\nodata & $H_\alpha$, $l'$, $r'$ imaging\tablenotemark{r} &  \cite{wkg+06} & No $H_\alpha$ excess \\
\enddata
\tablecomments{ Horizontal lines indicate the beginning and end of the active phase.}
\tablenotetext{a}{The Australia Telescope National Facility Parkes radio telescope}
\tablenotetext{b}{The Very Large Array, operated by the National Radio Astronomy Observatory}
\tablenotetext{c}{The SDSS 2.5 m telescope at the Apache Point Observatory}
\tablenotetext{d}{{Lick Observatory 3 m telescope}}
\tablenotetext{e}{ESO Faint Object Spectrograph and Camera on the 2.2 m telescope at La Silla}
\tablenotetext{f}{Kitt Peak National Observatory 2.1 m}
\tablenotetext{g}{Double-imaging spectrograph on the Apache Point Observatory 3.5 m telescope}
\tablenotetext{h}{Kitt Peak Bok Telescope, the spectropolarimeter SPOL}
\tablenotetext{i}{The 1.9 m telescope at the Sutherland site of the South African Astronomical Observatory}
\tablenotetext{j}{The 1.0 m telescope at the Sutherland site of the South African Astronomical Observatory}
\tablenotetext{k}{The 6.5 m Multi-Mirror Telescope}
\tablenotetext{l}{XMM-Newton EPIC-pn}
\tablenotetext{m}{XMM-Newton EPIC-MOS1/2}
\tablenotetext{n}{XMM-Newton OM}
\tablenotetext{o}{US Naval Observatory Flagstaff Station 1 m telescope}
\tablenotetext{p}{The Hiltner 2.4 m telescope}
\tablenotetext{q}{The 1.3 m McGraw-Hill telescope}
\tablenotetext{r}{The 2.5 m Isaac Newton Telescope}
\end{deluxetable}
\begin{deluxetable}{lccccl}
\tablecaption{\label{tab:ourobs} Our Observations of J1023.}
\tabletypesize{\scriptsize}
\tablehead{\colhead{Date} & \colhead{MJD range} & \colhead{Orbital phase} & \colhead{Instrument} & \colhead{Frequency (MHz)} & \colhead{Comment}}
\startdata
2007 Jun 25 & 54276.00 & 0.85 & GBT Spigot\tablenotemark{a} & 350 & Nearby track \\
2007 Jun 28 & 54279.98 & 0.96 & GBT Spigot\tablenotemark{a} & 350 & Discovery \\
2008 Oct 18 & 54757.54 & 0.71 & GBT Spigot\tablenotemark{\ast,a} & 350 & \\
2008 Oct 18 & 54757.58--54757.66 & 0.91--1.32 & GBT GUPPI\tablenotemark{\ast}/Spigot\tablenotemark{b} & 2000 & \\
2008 Oct 18 & 54757.68 & 0.40 & GBT GUPPI/Spigot\tablenotemark{b} & 2000 & \\
2008 Oct 18 & 54757.74 & 0.69 & GBT GUPPI/Spigot\tablenotemark{b} & 2000 & \\
2008 Oct 21 & 54760.76--54760.79 & 0.95--1.09 & GBT Spigot\tablenotemark{\ast,a} & 800 & \\
2008 Oct 22 & 54761.47--54761.51 & 0.57--0.73 & GBT Spigot\tablenotemark{\ast,a} & 800 & \\
2008 Oct 22 & 54761.53--54761.55 & 0.83--0.96 & AO ASP/WAPPs\tablenotemark{c} & 327 & \\
2008 Oct 24 & 54763.82--54764.06 & 0.39--1.59 & Parkes 10 cm\tablenotemark{\ast}/50 cm\tablenotemark{d} & 3000/700 & \\
2008 Oct 27  & 54766.53--54766.54 & 0.09--0.14 & AO ASP\tablenotemark{\ast}/WAPPs\tablenotemark{e} & 1400 & \\
2008 Nov 1 & 54771.50--54771.68 & 0.20--1.10 & GBT Spigot/GUPPI/GASP\tablenotemark{\ast,f} & 1400 & \\
2008 Nov 6  & 54776.49--54776.50 & 0.37--0.43 & AO ASP/WAPPs\tablenotemark{e} & 1400 & \\
2008 Nov 11 & 54781.03--54781.25 & 0.30--1.39 & WSRT PuMa-II\tablenotemark{g} & 150 & \\
2008 Nov 12 & 54782.47--54782.50 & 0.55--0.70 & AO ASP\tablenotemark{\ast}/WAPPs\tablenotemark{e} & 1400 & \\
2008 Nov 18 & 54788.45--54788.48 & 0.77--0.91 & AO ASP\tablenotemark{\ast}/WAPPs\tablenotemark{e} & 1400 & \\
2008 Nov 24 & 54794.43--54794.46 & 0.92--1.08 & AO ASP\tablenotemark{\ast}/WAPPs\tablenotemark{e} & 1400 & \\
2008 Dec 2 & 54802.42--54802.44 & 0.28--0.40 & AO ASP/WAPPs\tablenotemark{e} & 1400 & \\
2008 Dec 12 & 54812.39--54812.42 & 0.62--0.74 & AO ASP\tablenotemark{\ast}/WAPPs\tablenotemark{e} & 1400 & \\
2008 Dec 22 & 54822.36--54822.39 & 0.95--1.09 & AO WAPPs\tablenotemark{h} & 1400 & \\
2008 Dec 31 & 54831.09--54831.33 & 0.03--1.22 & WSRT PuMa-II\tablenotemark{i} & 350 & \\
2009 Jan 2 & 54833.33--54833.36 & 0.31--0.48 & AO ASP/WAPPs\tablenotemark{e} & 1400 & \\
2009 Jan 5 & 54836.16--54836.33 & 0.60--1.48 & GBT GASP\tablenotemark{\ast}/GUPPI\tablenotemark{j} & 1700 & \\
2009 Jan 16 & 54847.28--54847.30 & 0.76--0.84 & AO ASP\tablenotemark{\ast}/WAPPs\tablenotemark{e} & 1400 & \\
\enddata
\tablenotetext{\ast}{Used in the timing solution.}
\tablenotetext{a}{We used the Spigot with a 50 MHz band with 2048 frequency channels and a time resolution of $81.92~\mu\text{s}$.}
\tablenotetext{b}{We used the Spigot with an 800 MHz band with 2048 frequency channels and a time resolution of $81.92~\mu\text{s}$. GUPPI recorded full Stokes parameters for an 800 MHz band with 2048 frequency channels and a time resolution of $40.96~\mu\text{s}$. In both cases only about 600 MHz of the bandwidth is usable.}
\tablenotetext{c}{We used two WAPPs, each with a 12.5 MHz band with 512 frequency channels and a time resolution of $64~\mu\text{s}$. ASP recorded a bandwidth of 16 MHz centered at 327 MHz.}
\tablenotetext{d}{The dual-band receiver synthesized two filterbanks, one 64 MHz wide with 256 channels and centered at 590 MHz, and the other 768 MHz wide with 256 channels centered at 3032 MHz. Both had a time resolution of $80~\mu\text{s}$.}
\tablenotetext{e}{We used three WAPPs, with center frequencies 1170, 1370, and 1570 MHz, using bandwidths of 100 MHz with 256 channels and a time resolution of 64 $\mu$s. ASP recorded a 52 or 56 MHz bandpass centered on 1412 MHz.}
\tablenotetext{f}{We used the Spigot with an 800 MHz band with 2048 frequency channels and a time resolution of $81.92~\mu\text{s}$. GUPPI recorded full Stokes parameters for an 800 MHz band with 2048 frequency channels and a time resolution of $40.96~\mu\text{s}$. In both cases only about 600 MHz of the bandwidth is usable. GASP recorded an 84 MHz bandpass centered at 1388 MHz.}
\tablenotetext{g}{We used the PuMa-II backend to coherently dedisperse 8 2.5 MHz channels.}
\tablenotetext{h}{We used the same three-WAPP configuration as in f above, but the ASP data was accidentally corrupted.}
\tablenotetext{i}{We used the PuMa-II backend to coherently dedisperse 6 10 MHz channels.}
\tablenotetext{j}{We used GUPPI in an online folding mode, recording full Stokes parameters for an 800 MHz band with 2048  frequency channels and a time resolution of $40.96~\mu\text{s}$; only 600 MHz of the bandwidth was usable. GASP recorded a 76 MHz bandpass centered at 1788 MHz.}
\end{deluxetable}

\clearpage
\nocite{scilastnote}
\bibliography{journals,refs}{}

\begin{thebibliography}{10}
\bibitem{rochelobe}
 In a circular binary system, a star's Roche lobe is that volume in which
  matter corotating with the system is gravitationally bound to that star.
\end{thebibliography}
\bibliographystyle{Science}

\end{document}